\documentstyle[12pt]{article}
\input{psfig}
\begin{document}
\vspace*{-6ex}
\begin{flushright}
\fbox{RU 95/E-39}
\end{flushright}
\vglue 0.4cm
\begin{center}
{\bf RENORMALIZED DIFFRACTIVE CROSS SECTIONS AT HERA\\
AND THE STRUCTURE OF THE POMERON}
\footnote{Talk given at the ``VI$^{th}$
Blois Workshop, FRONTIERS IN STRONG INTERACTIONS"\\
Ch\^{a}teau de Blois, France, June 20-24, 1995.}
\vskip 1cm
K. GOULIANOS\\
{\em The Rockefeller University\\
1230 York Avenue, New York, NY 10021}\\
\end{center}
\vspace{0.75cm}

\begin{center}
\parbox{12.6cm}
{\begin{center} {\em ABSTRACT} \end{center}
\hspace*{0.3cm}
A phenomenological renormalization scheme for hadronic
diffraction is proposed, which achieves unitarization
without the need for
``screening corrections".  Predictions for diffractive
photoproduction cross sections at HERA are presented and compared with
experimental results. A new interpretation of hard and deep
inelastic diffractive
data
emerges, in which the momentum sum rule is obeyed by the constituents of
a pomeron described as a mixture of quark and gluon
color singlets in a ratio dictated by asymptopia.
}
\end{center}

\clearpage
\setlength{\baselineskip}{3.2ex}
\section{Introduction}

The early success of the concept of the
pomeron in describing elastic, diffractive, and total
cross sections in a simple Regge-pole model \cite{KG,DL1}
was tempered  by the measurements of $p\bar p$
single diffraction (SD) dissociation cross sections at the S$p\bar p$S Collider
\cite{UA4} and at the Tevatron \cite{E710,CDF}.  These measurements
showed that the rise of the SD cross section with energy
is too slow relative to that predicted by the theory.
Such a result was, of course, not
unexpected, since it was well known that the SD cross section in Regge theory
with a pomeron trajectory intercept  $\alpha(0)>1$ rises faster
than the total cross section, and if this rise continued it would lead
to violation of unitarity at the TeV energy scale.
However, the need for
{\em unitarizing} the simple Regge-pole description of cross sections
was elevated to a crisis by
the high energy SD measurements, and several schemes,
e.g. \cite{Shuler,GLM1,Kaidalov}{\hspace{1em},}  were
proposed to implement unitarization.
In this paper, we first discuss briefly a proposal  \cite{Renormalization} for
a phenomenological unitarization procedure
of hadronic diffraction
that gives the observed energy dependence
for the $pp/p\bar p$ single diffractive cross section, and then
we use this procedure to calculate photoproduction cross sections
at HERA energies.  Furthermore, by applying it to hard
and deep inelastic diffractive data, we obtain a picture in which the pomeron
consists of an asymptotic mixture of quark and gluon color singlets in a ratio
dictated by the quark counting rules for the asymptotic regime.
\section{Renormalization of hadronic diffraction}
The cross section for
single diffraction dissociation
in Regge theory has the form
\begin{equation}
\frac{d^2\sigma_{sd}^{ij}}{dtd\xi}=
\frac{1}{16\pi}\;\frac{\beta_{i{\cal{P}}}^2(t)}
{\xi^{2\alpha(t)-1}}\left[ \beta_{j{\cal{P}}}(0)\,g(t)
\;\left(\frac{s'}{s'_0}\right)^{\alpha(0)-1}\right]=f_{{\cal{P}}/i}(\xi,t)\;
\sigma_T^{{\cal{P}}j}(s',t)
\label{diffractive}
\end{equation}
where ${\cal{P}}$ stands for pomeron, $s'$ is the s-value in the
${\cal{P}}-j$ reference frame, $s'_0$ is a constant, $\xi=s'/s$ is the
{\mbox{Feynman-$x$}} of the pomeron in hadron-$i$, and $\alpha(t)$ the pomeron
trajectory given by
$\alpha(t)=\alpha(0)+\alpha't=1+\epsilon+\alpha't$.
\begin{figure}[h,t,b]
\vspace*{-0.35in}
\centerline{\psfig{figure=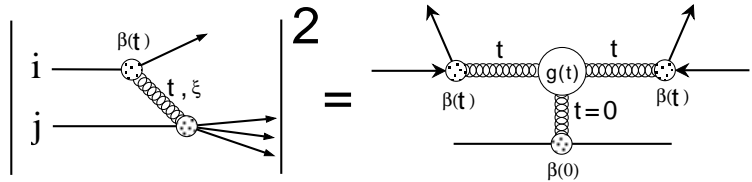,width=6in}}
\vspace*{-6.5in}
\caption{The triple-pomeron amplitude for single diffraction dissociation.}
\label{triple-pomeron}
\end{figure}
The term in the square brackets
is interpreted as being the pomeron-$j$ total cross section,
$\sigma_T^{{\cal{P}}j}(s',t)$, where $g(t)$ is the
``triple-pomeron coupling constant".
This interpretation leads naturally to viewing single diffraction
as being due to a flux of pomerons, $f_{{\cal{P}}/i}(\xi,t)$, {\em emitted}
by hadron-$i$ and interacting with hadron-$j$ (see Fig.~\ref{triple-pomeron}).

The term $\beta_{i{\cal{P}}}^2(t)$ in the pomeron flux factor
can be expressed in terms of the total cross section $\sigma_T^{ii}(s)$
and the elastic form factor $F^i(t)$ (obtained from
$d\sigma_{el}^{ii}/dt$):
\begin{equation}
\beta_{i{\cal{P}}}^2(t)=\beta_{i{\cal{P}}}^2(0)\;[F^i(t)]^2=
\sigma_T^{ii}(s)\;\left[\left(\frac{s}{s_0}\right)^
{\alpha(0)-1}\right]^{-1}\;[F^i(t)]^2
\end{equation}
The same expression can be used to write $\beta_{j{\cal{P}}}(0)$ in terms
of $\sigma_T^{jj}(s)$ and $s_0$, using $F^j(0)=1$.

The constants $s_0$ and $s'_0$, which represent energy scales
in the pomeron propagator, are
not specified by the theory, and experiment shows \cite{KG} that $g(t)$ is
independent of $t$,
i.e. $g(t)=g(0)$.  For a {\em universal} pomeron the energy scale
should be process independent and hence  $s'_0=s_0$.
Thus, there are
two free parameters in formula (\ref{diffractive}) for single diffraction,
$s_0$ and $g(0)$.
Since $s_0$ appears both in the flux factor
and in the ${\cal{P}}j$ cross section, and since
from the SD cross section as given in (\ref{diffractive})
one can determine only the product $\left[s_0^{\alpha(0)-1}\right]^
{1/2}g(0)$,
the normalization of the pomeron flux cannot be determined uniquely by the SD
data {\em in the standard Regge theory}. However, in the
{\em pomeron flux renormalization}
scheme that we propose, the constants
$s_0$ and $g(0)$ can be determined {\em independently} from SD data,
resulting in a uniquely normalized pomeron flux.

\begin{figure}[h,t,b]
\vspace*{-0.75in}
\centerline{\psfig{figure=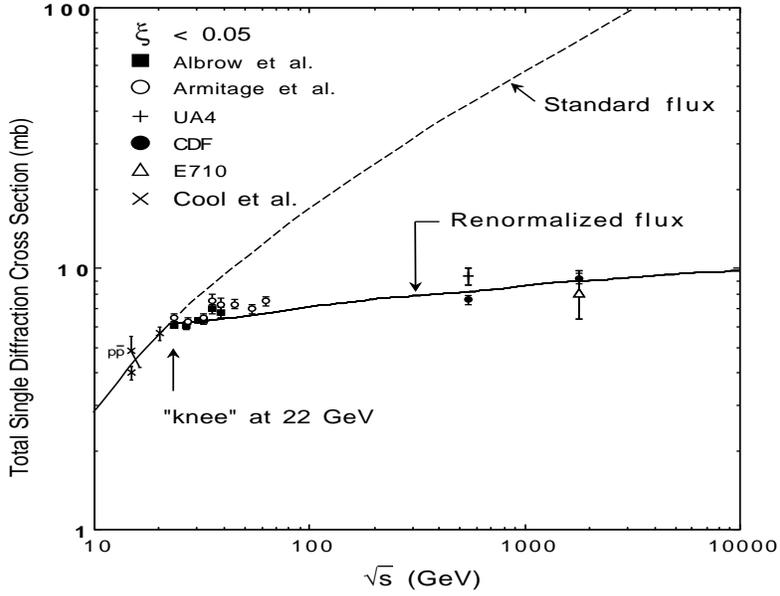,height=5in,width=5in}}
\vspace*{-1.25in}
\caption{SD cross section data
\protect{$2\sigma_T^{p(\bar p)-p}$}
for \protect{$\xi<0.05$} compared with predictions using the
standard pomeron flux (dashed line) and the renormalized flux (solid line).}
\label{SD}
\end{figure}

As mentioned in the introduction,
in the standard theory the  $p\bar p$ single diffractive
cross section rises much faster than that observed,
reaching the total cross section and therefore violating unitarity
at the TeV energy scale (dashed line in Fig.~\ref{SD}).  The
renormalization scheme that we
propose  treats the pomeron flux
as a probability density whose integral over $\xi$ and $t$ is not allowed
to exceed unity.  The renormalized flux is given by
\begin{equation}
f_N(\xi,t)=\frac{f_{{\cal{P}}/i}(\xi,t)}{N(\xi_{min})}
;\;\;\;\;N(\xi_{min})=\left\{ \begin{array}{l}
A(\xi_{min})\equiv
{\int_{\xi_{min}}^{0.1}d\xi \int_{t=0}^{\infty} f_{{\cal{P}}/i}(\xi,t)\;dt}\\
1,\;\;\;\;\;\;\;\mbox{if $A(\xi_{min})< 1$}
\end{array}\right.
\label{FN}
\end{equation}
where $\xi_{min}$=$(1.5\; \mbox{GeV}^2/s)$ for $p\bar p$ soft SD.
The solid line in Fig.~\ref{SD}
shows an `eyeball' fit to the data using this flux
(the obvious systematics in the data do not permit a proper $\chi^2$-fit).  The
position of the `knee' in this curve occurs at the $\sqrt{s}$-value at
which the flux integral becomes unity, which depends on the parameter $s_0$.
Therefore, $s_0$ {\em is determined from the position of this
`knee' in the data}.  In Fig.~\ref{SD}, the `knee' occurs
at $\sqrt{s}=22\;GeV$ for $s_0=1\;GeV^2$.
\section{Proton-(anti)proton diffractive cross sections}
The total, elastic, and renormalized
single diffractive, double diffractive and double
pomeron exchange cross sections are evaluated and compared with data in
Ref.~\cite{Renormalization}{\hspace{0.5em}}.
The following parameters were used in Eq.~(1)
to produce the solid curve in Fig.~\ref{SD}

$$
\begin{array}{llll}
\epsilon 			&= 	&0.115	& \\
\alpha '			&=	&0.26	&{\mbox{GeV}}^2 \\
\beta_{p{\cal{P}}}(0)		&=	&6.06	&{\mbox{GeV}}^{-1}\\
F^p(t)			&=	&e^{2.3t}& \\
g(t)=g(0)			&=	&1.1	&{\mbox{GeV}}^{-1}\\
\sigma_0^{{\cal{P}}p}\equiv \beta_{p{\cal{P}}}(0)\times g(t)
				&=	&2.6	&{\mbox{mb}}\\
s'_0=s_0				&=	&1	&{\mbox{GeV}}^2
\end{array}
$$
These parameters were determined\cite{Renormalization}
from data including the recent CDF
measurements\cite{CDF}.  The proton form factor, $F^p(t)$,
is valid for $|t|<\approx 0.1$ GeV$^2$; to incorporate larger $t$-values,
one may  set\cite{DL2} $F^p(t)=F_1(t)$,
the isoscalar form factor measured in electron-nucleon scattering:
\begin{equation}
F_1(t)=\frac{4m^2-2.8t}{4m^2-t}\left[\frac{1}{1-t/0.7}\right]^2
\label{F1}
\end{equation}
Noting that $s'=M^2=s\xi$,
the renormalized total SD cross section is given by
\begin{equation}
\sigma_{sd,N}(\xi_{min}<\xi<\xi_{max})
=\sigma_0^{{\cal{P}}p}\;
\left(\frac{s}{s'_0}\right)^{\epsilon}\;
\int^{\xi_{max}}_{\xi{min}} \int_{t=0}^{\infty}
\xi^{\epsilon}\,f_N(\xi,t)\,d\xi dt
=\sigma_0^{{\cal{P}}p}\;
\left(\frac{s}{s'_0}\right)^{\epsilon}\;\left<\xi^{\epsilon}\right>
\label{SDN}
\end{equation}
where $\xi_{min}=(1.5\;{\mbox{GeV}}^2)/s$. Since $\xi_{min}$ decreases with
increasing $s$, $\left<\xi^{\epsilon}\right>$ also decreases
and therefore $\sigma_{sd,N}$ increases at a rate {\em slower}
than $s^{\epsilon}$, i.e.
slower than the pomeron exchange component of the
$p\bar p$ total cross section given by $\sigma_T^{p\bar p}(s)=
\beta^2_{p{\cal{P}}}(0)\;(s/s_0)^{\epsilon}$, which
dominates at high energies.
Thus, as required by unitarity,
the renormalized SD cross section remains safely below the total cross section
at all energies.

Above $\sqrt{s}=22$ GeV, where the pomeron flux factor integral becomes unity,
the total renormalized SD cross section, calculated from Eq.~\ref{SDN}
and multiplied by 2,
has an approximately logarithmic $s$-dependence given by\cite{Renormalization}
\begin{equation}
\sigma_{SD}^{p\bar p}(s)_{\xi<0.05}=4.3+0.3\,\mbox{ln}s\;\;\;[s\mbox{ in
GeV}^2]
\label{SD-ln}
\end{equation}
The double diffraction (DD) dissociation cross section for
$p\bar p\rightarrow X_1X_2$ depends on $s,\;M_1$ and $M_2$.  Numerical
values for renormalized DD cross sections are presented in
Ref\cite{Renormalization}.{\hspace{0.5em}.} In the region
$30<\sqrt{s}<1800$ GeV, the total DD cross section for
$M_1^2,M_2^2>1.5$ GeV$^2$ and $M_1^2M_2^2<0.1(s\,s_0)$,
which represents the {\em coherence condition} for DD, can be approximated by
\begin{equation}
\sigma_{dd}^{p\bar p}(s)\left(\frac{M_1^2M_2^2}{s\,s_0}<0.1\right)
\approx 4.3-0.2\,\mbox{ln}s\;\;\;\;[30<\sqrt{s}<1800\mbox{ GeV}]
\label{DD-ln}
\end{equation}
Finally, the renormalized total double pomeron exchange cross section
for $p\bar p\rightarrow p\bar{p}X$ within 1~GeV$^2<M^2<0.01s$
is approximately constant
at the level of $\sim 50-70\;\mu$b\cite{Renormalization}
in the energy range from ISR to LHC.

\section{Photoproduction cross sections at HERA}
At HERA, where $\sim 28$ GeV electrons are brought into collision with
$\sim 800$ GeV protons ($\sqrt{s}\approx 300$ GeV),
photoproduction cross sections have been measured
at $\gamma p$ c.m.  energies $W\sim 200$ GeV
using low-$Q^2$ photons whose energy is tagged
by measuring the energy of the outgoing electron in
$e+p\rightarrow (e+\gamma)+p\rightarrow e+X$.
In this section, we calculate the total, the SD and the
DD photoproduction cross sections at HERA from the corresponding
$p\bar p$ cross sections
by simply replacing in the Regge formulas for $p\bar p$ the proton-pomeron
coupling constant, $\beta_{p{\cal{P}}}(0)$, by $\beta_{\gamma {\cal{P}}}(0)$.
The latter is obtained from the cross section for
$\gamma p\rightarrow Xp$ at $\sqrt{s}\sim14$ GeV measured in a fixed target
experiment\cite{E612}.
The calculated
cross sections are compared with H1\cite{H11,H112} and ZEUS \cite{ZEUS1}
measurements at HERA.
\subsection{Pomeron-photon coupling constant}
Using factorization, the ratio of the pomeron-photon to pomeron-proton
couplings, $\beta_{\gamma/p}$,
is equal to the ratio of the cross sections for $\gamma p\rightarrow Xp$ and
$pp\rightarrow Xp$:
\begin{equation}
\beta_{\gamma /p}\equiv
\frac{\beta_{\gamma {\cal{P}}}(0)}{\beta_{p{\cal{P}}}(0)}=
\frac{\sigma_{sd}^{\gamma p\rightarrow Xp}}{\sigma_{sd}^{pp\rightarrow Xp}}
=\frac{(6.59\pm0.32\pm 0.86)\,\mu\mbox{b}}{0.5\times4.3\mbox{ mb}}=
(3.06\pm 0.15\pm 0.4)\,10^{-3}
\label{gammaP}
\end{equation}
The SD cross sections here are for $\sqrt{s}=14$ GeV in the region
$M^2>1.5$ GeV$^2$ and $\xi<0.05$.
The value of $\sigma_{sd}^{\gamma p\rightarrow Xp}$
was obtained by integrating
the first term in Eq.~(9) of Ref.\cite{E612} using the parameters
provided in Table~1 of Ref.\cite{E612} (the second error is due to
the quoted $\pm13$\%
normalization uncertainty).
The value of $\sigma_{sd}^{pp\rightarrow Xp}$ was obtained
from Fig.~2.  Below, we use
$\beta_{\gamma /p}$ to predict photoproduction cross sections
for HERA.
\subsection{Total photoproduction cross section}
At $W\sim 200$ GeV, the total $\gamma p$ cross section
is completely dominated by pomeron exchange\cite{DL1} and therefore is
expected to be given by
\begin{equation}
\sigma_T^{\gamma p}(W)=\beta_{\gamma /p}\;\,\sigma_{T,\cal{P}}^{p\bar p}
(s=W^2)=
(3.06\pm0.15\pm 0.4)\times 10^{-3}\;
\left[\beta^2_{p\cal{P}}(0)\,\left(\frac{W}{W_0}\right)^{2\epsilon}\right]
\label{GPT}
\end{equation}
where $\sigma_{T,\cal{P}}^{p\bar p}$ is the portion of  $\sigma_T^{p\bar p}$
attributed to pomeron exchange\cite{Renormalization} and $W_0=1$ GeV.
Cross sections calculated using this formula are
compared below with data from H1\cite{H11,H112} and ZEUS\cite{ZEUS1}:

\begin{center}
\begin{tabular}{cccl}
$W$&$\sigma_T^{\gamma p}(predicted)$&$\sigma_T^{\gamma p}(measured)$
& \\
 & & & \\
197 GeV&$147.5\pm 7.2\pm 19\;\mu$b&$156\pm 2\pm 18\;\mu$b
&(H1, Ref.\cite{H11})\\
200 GeV&$148.0\pm 7.2\pm 19\;\mu$b&$165\pm 2\pm 11\;\mu$b&(H1, Ref.
\cite{H112})\\
180 GeV&$144.5\pm 7.1\pm 19\;\mu$b&$143\pm 4\pm 17\;\mu$b&(ZEUS)
\end{tabular}
\end{center}
Within the experimental errors, there is good agreement between the
predicted and measured total cross section values.
\subsection{Diffractive photoproduction cross sections}
The photoproduction cross sections for $\gamma p\rightarrow Xp$
and $\gamma p\rightarrow X_1X_2$ can be obtained by multiplying
the corresponding $pp$ cross sections by $\beta_{\gamma /p}$.
For $M_1^2, M_2^2>1.5$ GeV$^2$,
\begin{equation}
\sigma^{\gamma p\rightarrow Xp}(W)|_{\xi<0.05}
=\beta_{\gamma /p}\;\,
\frac{1}{2}\sigma_{sd}^{p\bar p}(s=W^2)|_{\xi<0.05}
=(3.06\,10^{-3})\;\,\frac{1}{2}
(4.3+0.3\,\mbox{ln}W^2)\;\mbox{mb}
\label{GSD}
\end{equation}
\begin{equation}
\sigma^{\gamma p\rightarrow X_1X_2}(W)
\left(\frac{M_1^2M_2^2}{W^2\,W^2_0}<0.1\right)
=\beta_{\gamma /p}\;\sigma_{dd}^{p\bar p}(s=W^2)\approx (3.06\,10^{-3})
\;(4.3-0.2\,\mbox{ln}W^2)\;\mbox{mb}
\label{GDD}
\end{equation}
At $W=200$ GeV, $\sigma^{\gamma p\rightarrow Xp}=(11.44\pm0.56\pm1.49)
\;\mu$b and
$\sigma^{\gamma p\rightarrow X_1X_2}=(6.7\pm 0.33\pm0.87)\;\mu$b.
Similar results have been obtained by calculating the ``screening corrections"
in an eikonal model\cite{GLM}.
We emphasize that the above cross sections are for a minimum diffractive
mass limit of
$M_{min}^2=1.5$ GeV$^2$.  In order to compare our
calculated SD
cross section with the recent measurement of $26\pm5\;\mu$b reported by H1
\cite{H112}, we must extend the low mass limit down to $M_{min}^2=m_{\rho}^2
+0.2=0.79$ GeV$^2$ and the upper mass limit to $M_{max}^2=0.1W^2$. This
extension of the mass limits increases our SD
cross sections by a factor of 1.2, yielding 7.9 $\pm0.4\pm 1.0\;\mu$b
for the experimental value at
$W=14$ GeV  and $13.7\pm0.7\pm1.8\;\mu$b for the calculated value
at $W=200$ GeV. The latter
is about half the cross section reported by H1.
Comparing now directly the experimental results of H1 and Ref.\cite{E612},
we note that the H1
single diffraction
cross section  at 200 GeV is 3.3 times larger than the cross section
at $W=14$ GeV, i.e. it scales as $(W^2)^{2\epsilon}$, where $\epsilon=0.115$.
Within the Regge framework, such an increase of the cross
section with energy  would occur only if
``screening corrections" were altogether absent, contrary to theoretical
expectations for substantial (factor of $\sim 2$) screening corrections
\cite{GLM,CKMT}.

\section{The structure of the pomeron}
The structure of the pomeron has been investigated in
$p\bar p$ colliders by UA8\cite{UA8}, which observed diffractive dijets
at $\sqrt{s}=630$ GeV and $|t|\sim 1.5$ GeV$^2$, and by CDF\cite{CDF2},
which searched for and placed upper limits for diffractive dijet
and $W$ production at $\sqrt{s}=1800$ GeV and $|t|\sim 0$.
At HERA, the quark content of the pomeron has been probed directly
with virtual high-$Q^2$ photons in $e^-p$ deep inelastic scattering.
Both the H1 \cite{H12} and ZEUS \cite{ZEUS2} Collaborations have
reported measurements of
the diffractive structure function $F^D_2(Q^2,\xi,\beta)$
(integrated over $t$, which is not measured),
where $\beta$
is the fraction of the pomeron's momentum carried by the quark being struck.
The experiments find
that the $\xi$-dependence factorizes out and has the form
$1/\xi^{1+2\epsilon}$, which is the same as the expression in the
pomeron flux factor (see Eq.~\ref{diffractive}).
Moreover, the fits yield $\epsilon\approx 0.1$, which is in agreement
with the value measured in {\em soft}
collisions.

The pomeron structure function was evaluated
from the H1 results in Ref.\cite{KG3} using the renormalized pomeron
flux. H1 integrates the diffractive form factor
$F^D_2(Q^2,\xi,\beta)$ over $\xi$ and provides values for the expression
\begin{equation}
\tilde{F}^D_2(Q^2,\beta)=\int_{0.0003}^{0.05}F^D_2(Q^2,\xi,\beta)d\xi
\label{F-tilde}
\end{equation}
The pomeron structure function is related to $\tilde{F}^D_2(Q^2,\beta)$ by
factorization:
\begin{equation}
\tilde{F}^D_2(Q^2,\beta)=\left[\frac{\int_{0.0003}^{0.05}d\xi \int_0^{\infty}
f_{{\cal{P}}/p}(\xi,t)\;dt}{N(s,Q^2,\beta)}\right]\,F_2^{{\cal{P}}}(Q^2,\beta)
\label{F2P}
\end{equation}
The expression in the brackets is the normalized flux factor.
For fixed $Q^2$ and $\beta$, $\xi_{min}=
(Q^2/\beta s)$.  Therefore, the flux integral, which to a good
approximation varies as $\xi_{min}^{-2\epsilon}$, is given by
\begin{equation}
N(\xi_{min})=
N(s,Q^2,\beta)\approx \left(\frac{\beta s}{Q^2}\;\xi_0\right)^{2\epsilon}=
3.8\left(\frac{\beta}{Q^2}\right)^{0.23}
\label{N}
\end{equation}
where $\xi_0$ is the value of $\xi_{min}$ for which the flux integral is unity.
For our numerical evaluations we use $\sqrt{s}$=300 GeV and the flux factor
of Ref.~\cite{Renormalization}, in which $\epsilon=0.115$.
The value of $\xi_0$ turns out to be $\xi_0=0.004$.

\begin{figure}[h,t,b]
\vspace*{-0.35in}
\centerline{\psfig{figure=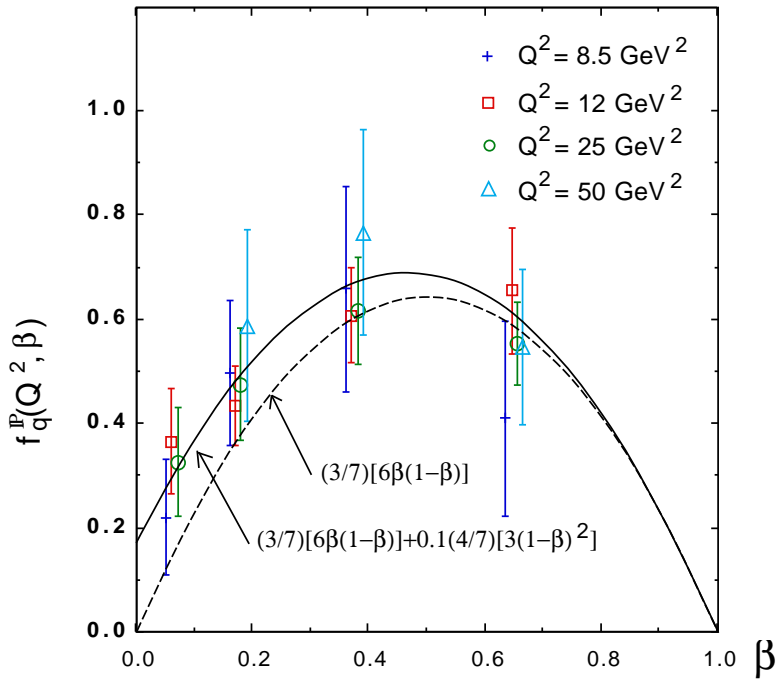,width=5in}}
\vspace*{-2.5in}
\caption{The quark compoment of the pomeron seen in DIS is compared to the
prediction (solid line) based on four quark flavors
and a pomeron that obeys the momentum sum rule; the dashed line represents
the direct quark contribution.}
\label{F2}
\end{figure}

Assuming now that the pomeron structure function receives contributions
from the four lightest quarks, whose average charge squared is 5/18, the
quark content of the pomeron is
\begin{equation}
f^{{\cal{P}}}_{q}(Q^2,\beta)=\frac{18}{5}\,F_2^{{\cal{P}}}(Q^2,\beta)
\label{FQ}
\end{equation}
The values of $f^{{\cal{P}}}_q(Q^2,\beta)$ obtained in this manner are
shown in Fig.~\ref{F2}.

As seen, the renormalized points show no $Q^2$
dependence. We take this fact as an indication that
the pomeron ``lives" in the asymptotic regime
and compare the data points with the asymptotic
momentum fractions expected for any quark-gluon construct
by leading-order perturbative QCD, which for
$n_f$ quark flavors are
\begin{equation}
f_q=\frac{3n_f}{16+3n_f}\hspace*{0.5in} f_g=\frac{16}{16+3n_f}
\label{flavors}
\end{equation}
For $n_f=4$, $f_q=3/7$ and $f_g=4/7$.
The quark and gluon components of the pomeron structure
are taken to be
 $f_{q,g}^{{\cal{P}}}(\beta)=f_{q,g}\;[6\beta(1-\beta)]$.
The pomeron in this picture is
a combination of valence quark and gluon color singlets and its complete
structure function, which obeys the momentum sum rule, is given by
\begin{equation}
f^{{\cal{P}}}(\beta)=\frac{3}{7}[6\beta(1-\beta)]_q+
\frac{4}{7}[6\beta(1-\beta)]_g
\label{fP}
\end{equation}
The data in Fig.~\ref{F2} are in
reasonably good
agreement with the quark-fraction of the structure function given by
$f_{q}^{{\cal{P}}}(\beta)=(3/7)[6\beta(1-\beta)]$, except for
a small excess at the low-$\beta$ region.  An excess at low-$\beta$
is expected in this
picture to arise from
interactions of the photon
with the gluonic part of the pomeron through gluon splitting into
$q\bar q$ pairs.  Such interactions, which are suppressed  by an order
of $\alpha_s$, result in an {\em effective} quark $\beta$-distribution
of the form $3(1-\beta)^2$. We therefore compare in Fig.~\ref{F2}
the data with the
distribution
\begin{equation}
f_{q,eff}^{{\cal{P}}}(\beta)=(3/7)[6\beta(1-\beta)]+\alpha_s(4/7)[3(1-\beta)^2]
\label{fqeff}
\end{equation}
using $\alpha_s=0.1$. Considering that this distribution involves
{\em no free parameters}, the agreement with the data is remarkable!
The UA8 and CDF results, and the recently reported \cite{ZEUS3} value
of (30-80)\%
for the hard-gluon content of the pomeron, are all consistent with this
picture.

\end{document}